\documentclass[aps,prl,preprint,superscriptaddress]{revtex4-1}
\usepackage{graphicx}
\usepackage{dcolumn}
\usepackage{bm}
\usepackage[usenames]{color}
\usepackage{amsmath}
\usepackage{amssymb}
\usepackage{latexsym}

\begin{document}

\title{Coexistence of the Kondo effect and spin glass physics in Fe-doped NbS$_2$}

\author{H Nobukane}
\affiliation{Department of Physics, Hokkaido University, Sapporo, 060-0810, Japan}
\affiliation{Center of Education and Research for Topological Science and Technology, Hokkaido University, Sapporo, 060-8628, Japan}

\author{Y Tabata}
\affiliation{Department of Physics, Hokkaido University, Sapporo, 060-0810, Japan}

\author{T Kurosawa}
\affiliation{Department of Physics, Hokkaido University, Sapporo, 060-0810, Japan}
\affiliation{Center of Education and Research for Topological Science and Technology, Hokkaido University, Sapporo, 060-8628, Japan}

\author{D Sakabe}
\affiliation{Department of Applied Physics, Hokkaido University, Sapporo, 060-8628, Japan}

\author{S Tanda}
\affiliation{Center of Education and Research for Topological Science and Technology, Hokkaido University, Sapporo, 060-8628, Japan}
\affiliation{Department of Applied Physics, Hokkaido University, Sapporo, 060-8628, Japan}

\date{\today}

\begin{abstract}
We report the coexistence of the Kondo effect and spin glass behavior in Fe-doped NbS$_2$ single crystals.
The Fe$_x$NbS$_2$ shows the resistance minimum and negative magnetoresistance due to the Kondo effect, and exhibits no superconducting behavior at low temperatures.
The resistance curve  follows a numerical renormalization-group theory using the Kondo temperature $T_K =12.3$~K for $x=0.01$ as evidence of Kondo effect.
Scanning tunneling microscope/spectroscopy (STM/STS) revealed the presence of Fe atoms near sulfur atoms and asymmetric spectra.
The magnetic susceptibility exhibits a feature of spin glass.
The static critical exponents determined by the universal scaling of the nonlinear part of the susceptibility suggest a three-dimensional Heisenberg spin glass.
The doped-Fe atoms in the intra- and inter-layers revealed by the X-ray result can realize the coexistence of the Kondo effect and spin glass.

\vspace{1cm}
\noindent
\textbf{Keywords:} Kondo effect, spin glass, transition metal dichalcogenides

\noindent
\textbf{E-mail:} nobukane@sci.hokudai.ac.jp
\end{abstract}

\maketitle

\subsection*{Introduction}

The research for transition metal dichalcogenides \textit{MX$_2$} (where \textit{M} is a group-4 or -5 metal and \textit{X} = S, Se or Te) has been revitalized by the discovery of a quantum Hall effect~\cite{Novoselov} and unconventional superconductivity in graphene~\cite{Herrero}.
Extending graphene research, novel emergent physical phenomena and applications can be investigated in \textit{MX$_2$}~\cite{Xi_NbSe2,Efetov}.
In particular, the relationship between a charge density wave (CDW) and superconducting states at low temperatures has generated interest in \textit{MX$_2$} systems for decades in terms of their competition or coexistence~\cite{Naito}.
The presence of a CDW in cuprate superconductors has been a recent topic~\cite{cuprate1,cuprate2}.
For Fe-based superconductors, the coexistence and competition of a spin-density-wave (SDW) and superconductivity has been debated~\cite{Wen}.
Thus, the studies of \textit{MX$_2$} play an important role in clarifying two-dimensional quantum physics.

The novel quantum states in \textit{MX$_2$} appear by controlling the carriers of materials, doping impurity and atomically thin layer thickness~\cite{Yoshida,Fe-TaS2,Xi_NbSe2,Zant_TaS2}.
For instance, the superconducting transition temperature in Cu$_x$TaS$_2$ first increases to 4.5~K at the optimal composition $x=0.04$, and then decreases at higher $x$~\cite{Cu-TaS2}, as observed with electron or hole doping in cuprate superconductors.
An atomically thin layer of \textit{MX$_2$} can control dimensionality without disorder.
The CDW transition temperature of NbSe$_2$ thin films increases from 33~K in bulk to 145~K, while the superconducting transition temperature decreases from 7.2~K in bulk to about 5~K with reducing the sample thickness \cite{Xi_CDW}.
The modulations of the transition temperature in ultrathin films are considered due to the effect of negative pressure in layered materials~\cite{Xi_CDW,Goli}.

2H-NbS$_2$ exhibits a peculiar difference from the properties of other 2H-\textit{MX$_2$} materials, such as 2H-NbSe$_2$, 2H-TaS$_2$ and 2H-TaSe$_2$~\cite{Neto}.
NbS$_2$ exhibits a superconductivity of $\sim 5$~K without showing CDW at higher temperature~\cite{Onabe,Suderow}.
Intriguingly, recent diffuse X-ray scattering measurements revealed traces of CDW in NbS$_2$~\cite{Leroux}.
Moreover it has been theoretically predicted that CDW and SDW instability is stabilized in Fe-doped monolayer for NbS$_2$~\cite{NbS2_SDW,Loon}.
Hence, we focus on the pairing mechanism of the superconductivity and the possibility of CDW formation in NbS$_2$.
The interplay between magnetism, Kondo effect and superconductivity can also be expected by doping magnetic atoms.
However, there have been few experimental studies of Fe-doped NbS$_2$.
To reveal the novel phenomena, we performed electrical and magnetic measurements on Fe$_x$NbS$_2$.

Here we report the electric and magnetic properties of Fe-doped NbS$_2$. The upturn resistance due to the Kondo effect was observed.
Scanning tunneling microscope/spectroscopy (STM/STS) results show Fe atoms near sulfur atoms and asymmetric spectra.
Magnetic measurements revealed the behavior of the spin glass.
From the results, we suggest that the Kondo effect and spin glass behavior coexist in layered Fe$_x$NbS$_2$.

\subsection*{Experimental}

Single crystals of Fe$_x$NbS$_2$ were grown by the chemical vapor transport method. The maximum size of the measured single crystals was about 3 mm~$\times$ 3 mm~$\times$ 0.4 mm.
We determined the lattice constants and polytypes by using X-ray diffraction measurement.
The electrical resistance was measured by the 4-terminal method at several Fe concentrations ($x=0.01, 0.02, 0.04~\mathrm{and}~0.08$) using the Physical Property Measurement System (PPMS) (Quantum Design) at $T$ down to 2~K with magnetic fields up to 14~T.
We performed the STM/STS measurement for $x=0$ and 0.01 samples at 8~K.
The crystals were cleaved in situ at 8~K.
In addition, we carried out a magnetic susceptibility measurement using the Magnetic Property Measurement System (MPMS) (Quantum Design).

\subsection*{Results and discussion}
\begin{figure}[t]
\begin{center}
\includegraphics[width=0.8\linewidth]{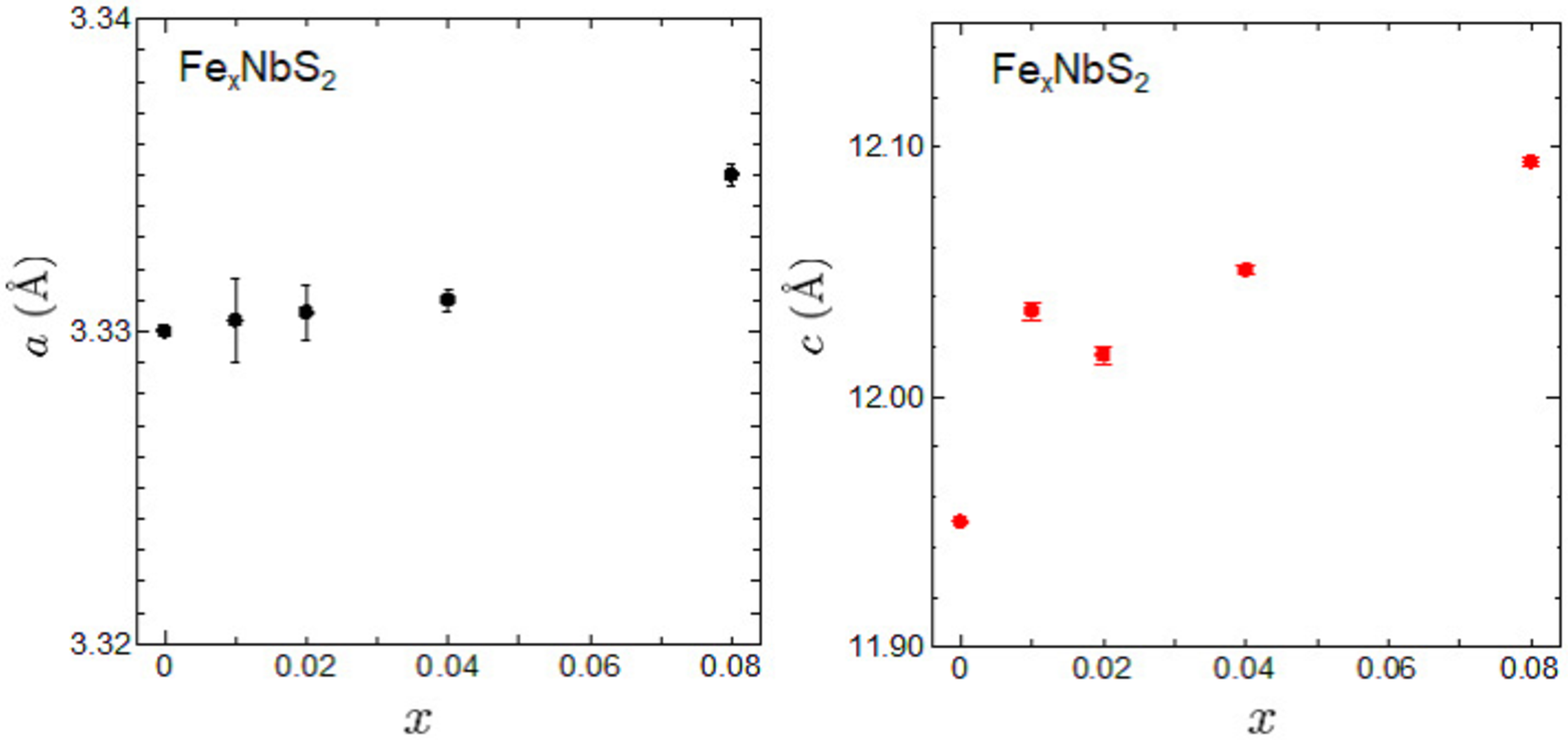}
\caption{
Fe concentration dependence of the lattice constants of the $a$ and $c$ axes for Fe$_{x}$NbS$_2$ at room temperature measured by X-ray diffraction.
}
\label{fig:x-ray}
\end{center}
\end{figure}

\begin{figure}[t]
\begin{center}
\includegraphics[width=0.8\linewidth]{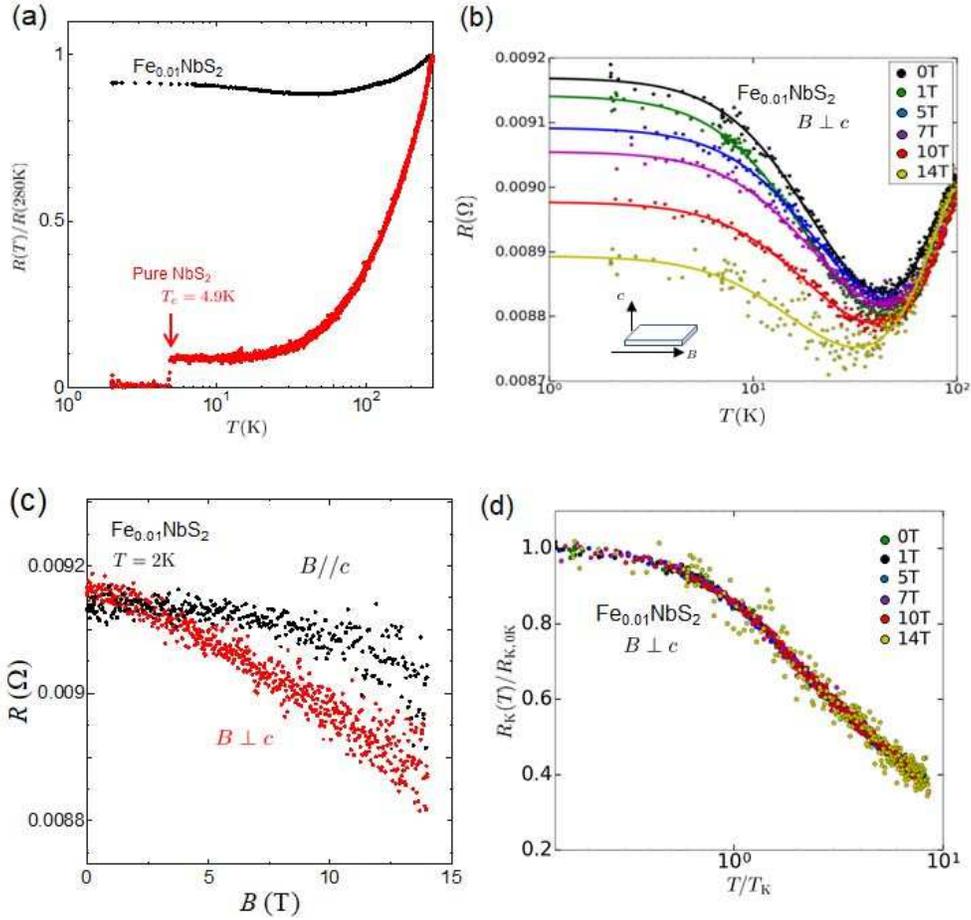}
\caption{Electric transport properties in Fe$_x$NbS$_2$.
(a)Temperature dependence of normalized resistance $R(T)/R(280K)$ for pure ($x=0$) and Fe$_{x=0.01}$NbS$_2$.
(b)Temperature dependence of resistance below 100~K for several magnetic fields. The solid curves represent the fitting result obtained with the Kondo model Eq.~(1-3).
(c)Magnetoresistance at 2~K in Fe$_{0.01}$NbS$_2$ for applied magnetic fields parallel and perpendicular to the $c$ axis.
(d)Universal Kondo behavior of normalized resistance versus temperature scaled to the Kondo temperature $T_{K}$ for several magnetic fields.}
\label{fig:kondo}
\end{center}
\end{figure}

Figure~\ref{fig:x-ray} shows the Fe concentration dependence of the lattice constants for Fe$_x$NbS$_2$.
The X-ray result shows that an increase of the $a$-axis of 0.15 \% and an increase of the $c$-axis of 1.5 \% between $x=0$ and $x=0.08$.
This indicates that Fe atoms mainly occupy inter-layer positions.
Figure~\ref{fig:kondo}(a) shows the temperature dependence of normalized resistance at 280~K for Fe$_x$NbS$_2$ at $x=0$ and $x=0.01$, respectively.
The superconducting transition temperature of pure NbS$_2$ is 4.9~K, which is consistent with previous reports~\cite{Naito,Onabe,Suderow}.
In contrast, with decreasing temperature, the resistance in $x=0.01$ samples exhibits metallic behavior from room temperature to about 50~K and upturns, and then the resistance saturates below about 10~K.
Figure~\ref{fig:kondo}(b) shows the temperature dependence of the resistance of $x=0.01$ for an applied magnetic field parallel to the $c$ axis.
A negative magnetoresistance was observed at lower temperature.
The change in the negative magnetoresistance for $B\perp c$ was larger than that for $B // c$ as shown in Fig.~\ref{fig:kondo}(c).
This behavior of the negative magnetoresistance is opposite to that for Anderson localization.
With Anderson localization, the change rate of the negative magnetoresistance becomes greater as the magnetic field is applied perpendicular to the conducting plane~\cite{Kawaji}.
For these reasons, we assumed that the upturn resistance behavior of Fe-doped samples is induced by localized Fe spin.
Here we used the Kondo model to analyze the resistance data of Fe$_x$NbS$_2$, which is given by Eq.~(1-3).

\begin{equation}
       R=R_0 + qT^2 + pT^5 + R_K(T/T_K)
\label{eq:Kondo}
\end{equation}
where $R_0$ is residual resistance due to sample disorder and $T^2$, $T^5$ and $R_K(T/T_K)$ represent the contributions of electron-electron scattering, electron-phonon scattering and the Kondo effect determined from a numerical renormalization group, respectively.

\begin{equation}
R_K\left(\frac{T}{T_K}\right)=R_{\mathrm{0K}}\left(\frac{{T^{\prime}}^2_K}{T^2 + {T^{\prime}}^2_K}\right)^s
\label{eq:Kondo2}
\end{equation}

\begin{equation}
       T^{\prime}_K=\frac{T_K}{(2^{1/s}-1)^{1/2}}
\label{eq:Kondo3}
\end{equation}
where $T_K$ is the Kondo temperature, and $s=0.225$ for a spin $1/2$ object is obtained from the numerical renormalization-group theory (NRG)~\cite{Costi,Gordon}.
In Fig.~\ref{fig:kondo}(b), solid lines represent the fitted curves obtained by using Eq.~(1-3).
A numerical fit in zero magnetic field yielded $R_0=8.1\times10^{-3}~\Omega$, $q=6.8 \times 10^{-8}~\Omega/\mathrm{K}^2$, $p=1.9 \times 10^{-15}~\Omega/\mathrm{K}^5$, $R_K(0~\mathrm{K})=1.1\times 10^{-3}~\Omega$, and $T_{K}=12.3~\mathrm{K}$.
Using the fitting parameters, we can scale the $R(T)$ curves at various magnetic fields in comparison with the universal Kondo behavior~\cite{Costi,Gordon,Han}.
Figure~\ref{fig:kondo}(d) shows the plot of normalized resistance $R_K(T)$/$R_{\mathrm{0K}}$ versus $T/T_K$ at each magnetic field.
The experimental data are fitted well to the universal Kondo behavior.
Thus, we emphasize that the upturns of the resistance are dominated by the Kondo effect.
We note that all the Fe$_x$NbS$_2$ samples with $x=0.01, 0.02, 0.04$ and $0.08$ exhibited the upturn resistance behavior of the Kondo effect and were scaled to universal curves of the NRG (see Supplementary Material, Fig.~S1).

\begin{figure}[t]
\begin{center}
\includegraphics[width=0.8\linewidth]{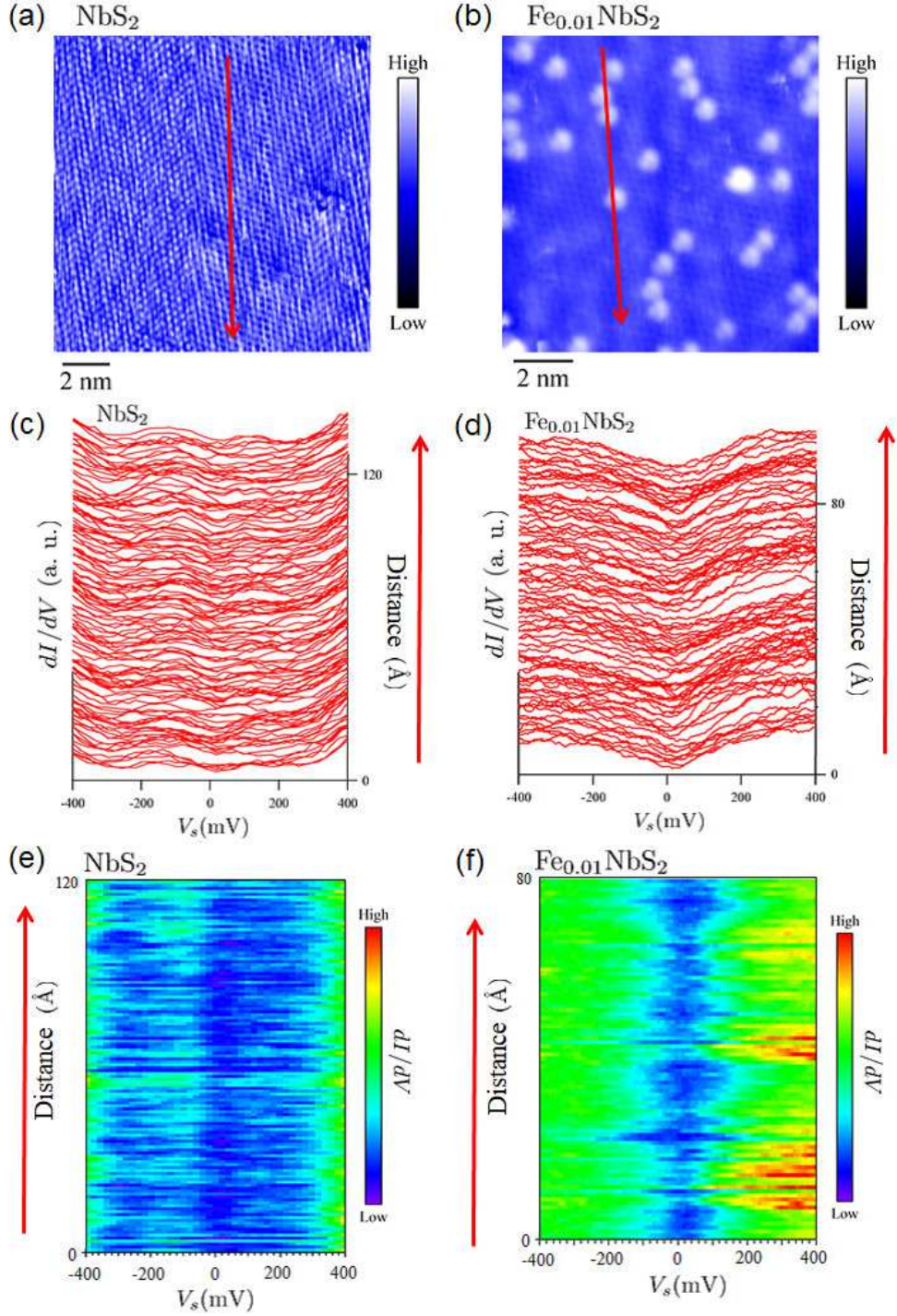}
\caption{STM images for (a)NbS$_2$ measured at $V_{s}=600$~mV and $I_{t}=0.8$~nA and (b)Fe$_{0.01}$NbS$_2$ measured at $V_{s}=-600$~mV and $I_{t}=0.48$~nA.
(c)-(d) Spatial dependence of STS spectra along the red arrows in STM images for Fe$_{x}$NbS$_2$ with $x=0$ (a) and $x=0.01$ (b).
(e)-(f) The color map of the spatial dependence of STS spectra.
The STS spectra around Fe atoms are asymmetric with respect to the bias voltage.
}
\label{fig:STM}
\end{center}
\end{figure}

Next, we performed STM/STS measurements to explore the local electric states around Fe atoms.
Figure \ref{fig:STM} shows STM/STS data on cleaved surfaces for $x=0$ and 0.01 samples, respectively.
In the STM image for $x=0$ (Fig.~\ref{fig:STM}(a)), the bright points correspond to sulfur atoms.
On the other hand, with the $x=0.01$ sample, large bright points can be seen that correspond to Fe atoms in addition to sulfur atoms as shown in Fig. \ref{fig:STM}(b).
This large bright spots correspond to a size of about three S atoms (small white points).
This suggests that dopant Fe atoms modulate the density of states when they are near S sites~\cite{Fan, Shu}.
Figures \ref{fig:STM}(c) and \ref{fig:STM}(d) show the spatial evolution of STS spectra along the red arrow in STM images for Fe$_x$NbS$_2$ with $x=0$ (Fig.~\ref{fig:STM}(a)) and $x=0.01$ (Fig.~\ref{fig:STM}(b)), respectively.
In pure NbS$_2$ ($x=0$), the STS spectra become symmetric as a function of bias voltage and are spatially homogeneous as shown in Figs.~\ref{fig:STM}(c) and~\ref{fig:STM}(e).
This means that the metallic state is realized at the measurement temperature of 8~K.
Interestingly, the STS spectra for $x=0.01$ exhibit spatial inhomogeneity around Fe atoms (large bright points), which is asymmetric with respect to the bias voltage as shown in Figs.~\ref{fig:STM}(d) and~\ref{fig:STM}(f) (See Supplementary Material, Fig. S2).
The asymmetric spectra suggest a Kondo resonance state measured by STM in the vicinity of the Fe atom~\cite{Madhavan,Li}.
The STM/STS results show the Kondo effect nearby Fe atoms in the intralayer.

\begin{figure}[t]
\begin{center}

  \includegraphics[width=0.9\linewidth]{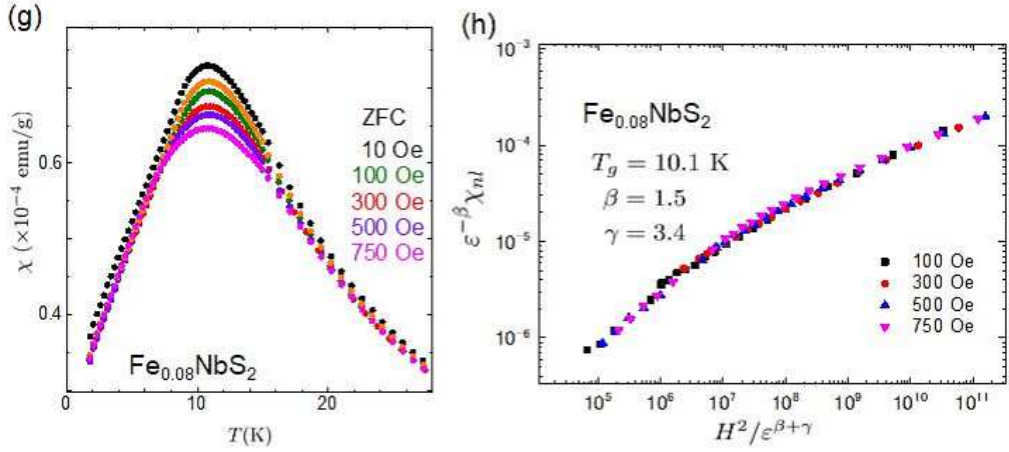}

\caption{Magnetic properties in Fe$_{x}$NbS$_2$ for $x=0.01$, 0.02, 0.04 and 0.08.
(a)Magnetic susceptibility $\chi$ of Fe$_{0.01}$NbS$_2$ as a function of temperature for various magnetic fields in the ZFC process.
The inset shows temperature dependence of ZFC and FC magnetization below 10~K.
(b)Scaling plot for Fe$_{0.01}$NbS$_2$ with $T_{g}=4.7~\mathrm{K}, \beta = 1.0$ and $\gamma =2.2$ as the nonlinear susceptibility $\chi_{nl}$ in various magnetic fields.
(c), (e) and (g)Temperature dependence of magnetic susceptibility for $x=0.02, 0.04$ and 0.08.
(d), (f) and (h)Scaling plot of the nonlinear susceptibility  according to Eq.~\ref{eq:chi_nl} for $x=0.02, 0.04$ and 0.08.
}
\label{fig:glass}
\end{center}
\end{figure}

To study the magnetic properties of Fe$_x$NbS$_2$, we measured the temperature dependence of zero field cooling (ZFC) and field cooling (FC) magnetization in a magnetic field of 1000~Oe as shown in the inset of Fig.~\ref{fig:glass}(a).
Above 5~K, the temperature dependence of magnetization can be described by the Currie-Weiss law, $\chi(T)=\chi_{0}+C/(T-\theta)$.
The best fit of the FC $\chi(T)$ data gives a negative effective $\theta=-4.4$ K, indicating the presence of the antiferromagnetic (AF) correlation.
Figure~\ref{fig:glass}(a) shows magnetic susceptibility as a function of temperature for various magnetic fields in the ZFC process.
We found a cusp at 4.7~K in the magnetic field of 10~Oe.
$\chi(T)$ deviates from the Currie-Weiss function toward smaller values in larger magnetic fields.
Figures~\ref{fig:glass}(c), \ref{fig:glass}(e) and \ref{fig:glass}(g) show temperature dependence of magnetic susceptibility for $x=0.02, 0.04$ and 0.08.
These behaviors exhibit a feature of a typical spin glass.
To analyze the spin glass transition, we carried out a static scaling analysis within the framework of the second-order phase-transition theory using spin correlation function ${M_{nl}}/H\propto{\sum_{i,}}_j[{\langle S_i,S_j\rangle}^{2}_{T}]$~\cite{Fischer,Sherriongton}.
The nonlinear susceptibility $M_{nl}/H(\equiv\chi_{nl})$ is defined as $\chi_{nl}(T, H)=\chi_{l}-\frac{M(T, H)}{H}$.
Here we assume that $\chi(T)$ measured at 10 Oe is a good approximation of the linear susceptibility $\chi_{l}$ above the spin-glass freezing temperature $T_g$, which is well described by the Currie-Weiss law.
For $\chi_{nl}$, the scaling theory predicts the relationship

\begin{equation}
\chi_{nl}\varepsilon^{- \beta}=F\left( \frac{H^2}{\varepsilon^{\beta + \gamma}}\right),
\label{eq:chi_nl}
\end{equation}
where $\varepsilon$ is the reduced temperature, $\varepsilon =(T-T_{g})/T_{g}$,  $\beta$ and $\gamma$  are the critical exponents, and $F$ is scaling function.
We found that the experimental data fell on a universal curve as shown in Fig.~\ref{fig:glass}(b).
For Fe$_{0.01}$NbS$_{2}$, we determined the critical exponents $\beta=1.0$ and $\gamma=2.2$ using our scaling analysis.
In addition, we obtained the critical exponents for $x=0.02$, 0.04, and 0.08 as shown in Figs.~\ref{fig:glass}(d), \ref{fig:glass}(f) and \ref{fig:glass}(h).
The results are shown in Table~\ref{table1}.
The static critical exponents for $x=0.01, 0.02$, and 0.04 are comparable to $\beta=1.0 \pm 0.1$ and $\gamma=2.2 \pm 0.1$ of three-dimensional Heisenberg-like spin glass~\cite{Bouchiat,Levy,Malinowski}.
The critical exponents for $x=0.08$ exhibits $\beta=1.5$ and $\gamma=3.4$, which is close to that of three-dimensional weakly anisotropic Heisenberg spin glass~\cite{Beauvillain,Neacsa}.
With increasing the Fe concentration, our results may suggest a crossover from the three-dimensional Heisenberg spin glass to the weakly anisotropic Heisenberg spin glass.

\begin{table}[h]
\begin{tabular}{lcc}
\hline\hline
$x$ & $\beta$ ~~~& $\gamma$ \\

\hline

0.01 \hspace{0.8cm}  & 1.0 ~~~ & 2.2 \\

0.02 & 1.0 ~~~& 2.1 \\

0.04 & 1.0 ~~~& 2.3 \\

0.08 & 1.5 ~~~& 3.4 \\

\hline\hline
\end{tabular}

\caption{The critical exponents measured for Fe$_x$NbS$_2$.}
\label{table1}
\end{table}

\begin{figure}[t]
\begin{center}
\includegraphics[width=0.8\linewidth]{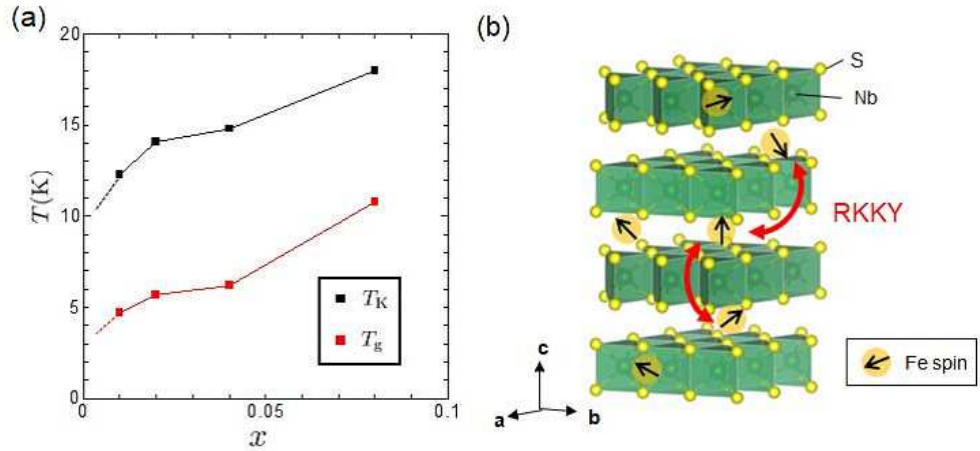}
\caption{(a)Phase diagram of $T_K$ and $T_g$ vs Fe concentration obtained from experimental data.
$T_K$ at zero magnetic field is plotted.
(b)Schematic of the coexistence of the Kondo effect and the spin glass in layered Fe$_x$NbS$_2$.
Localized spins in the conduction plane form the Kondo singlet state with conduction electrons.
Three-dimensional Heisenberg spin glass is appeared by Fe spins in the interlayer.
}
\label{fig:diagram}
\end{center}
\end{figure}

We revealed the Kondo effect and the spin glass from the electric transport and magnetic measurement results, respectively.
Figure~\ref{fig:diagram}(a) shows $T_{K}$ and $T_{g}$ plots as a function of Fe concentration.
The $T_{K}$ obtained from the fit for resistance data in zero magnetic field is plotted.
In general, the Kondo effect and the spin glass have a competitive relationship in conventional three-dimensional magnetic materials.
The Doniach phase diagram~\cite{Doniach} is described in terms of the competition between indirect local moment exchange, represented by the Ruderman-Kittel-Kasuya-Yoshida (RKKY) interaction $T_{RKKY}(T_{g}) \propto J^2$, and Kondo screening, characterized by $T_{K} \propto \mathrm{exp}(-1/J)$, where $J$ represents the effective spin-exchange coupling constant.
When $J$ is small, the RKKY exchange is dominant and in contrast the Kondo screening appears when $J$ is large.
Here, the Fe atom concentration in Fig.~\ref{fig:diagram}(a) corresponds to the exchange constant $J$.
The cerium and ytterbium compounds realize the competition between the Kondo effect and the spin glass with the AF correlation~\cite{Coqblin_Philo}.
Interestingly, $T_{g}$ and $T_{K}$ with $x$ dependence do not intersect in Fig.~\ref{fig:diagram}(a).
This means that the Kondo effect and the spin glass are not competitive but coexistent.
The coexistence of the Kondo effect and ferromagnetic order has been reported in uranium compounds~\cite{Bukowski,Tran}, which is described in the extended ferromagnetic Doniach phase diagram using a underscreened Kondo lattice model~\cite{Coqblin_MMM}.
We discuss the origin of the coexistence of the Kondo effect and the spin glass in Fe$_x$NbS$_2$ with the AF correlation.
We focus on where the doped Fe atoms are located in layered NbS$_2$ single crystals.
The X-ray result in Fig.~\ref{fig:x-ray} indicates that the doped Fe atoms exist in both the inter and intra-layers.
We suggest that the Kondo effect is induced by Fe atoms in a two-dimensional conduction plane.
Moreover, three-dimensional Heisenberg spin glasses are occurred by Fe atoms in the interlayer as shown in Fig.~\ref{fig:diagram}(b).
Thus, we conclude that the Kondo effect and the spin glass can coexist in two-dimensional layered NbS$_2$.
Here we describe why we used only resistance data in the discussion about the Kondo physics.
In general, the Kondo behavior can be discussed not only from resistance data but also from susceptibility data. Since our results suggest the coexistence of the Kondo effect and spin glass behavior, we think that the spin glass obscures the Kondo behavior in susceptibility.

In conclusion, we have reported the electric and magnetic properties of Fe-doped NbS$_2$.
When doped with Fe, Fe$_x$NbS$_2$ does not display a superconducting transition at low temperatures, and exhibits a saturating resistance upturn and negative magnetoresistance.
Asymmetric spectra were observed as a feature of the modulation of the electric density of states in the Fe-doped NbS$_2$.
The obtained static critical exponents suggest the presence of three-dimensional Heisenberg spin glasses.
We clarified that the Kondo effect and the spin glass can coexist due to the presence of Fe atoms in both intra- and inter-layers of NbS$_2$.

\subsection*{Acknowledgements}

We thank K. Isono, Y. Ogasawara, N. Matsunaga, S. Noro and K. Nomura for experimental help and useful discussions.
This work was supported by JSPS KAKENHI (No. 17K14326), Inamori Foundation, Iketani Science and Technology Foundation and Samco Science and Technology Foundation.

\newpage

\begin{figure}[t]
\begin{center}
\includegraphics[width=0.9\linewidth]{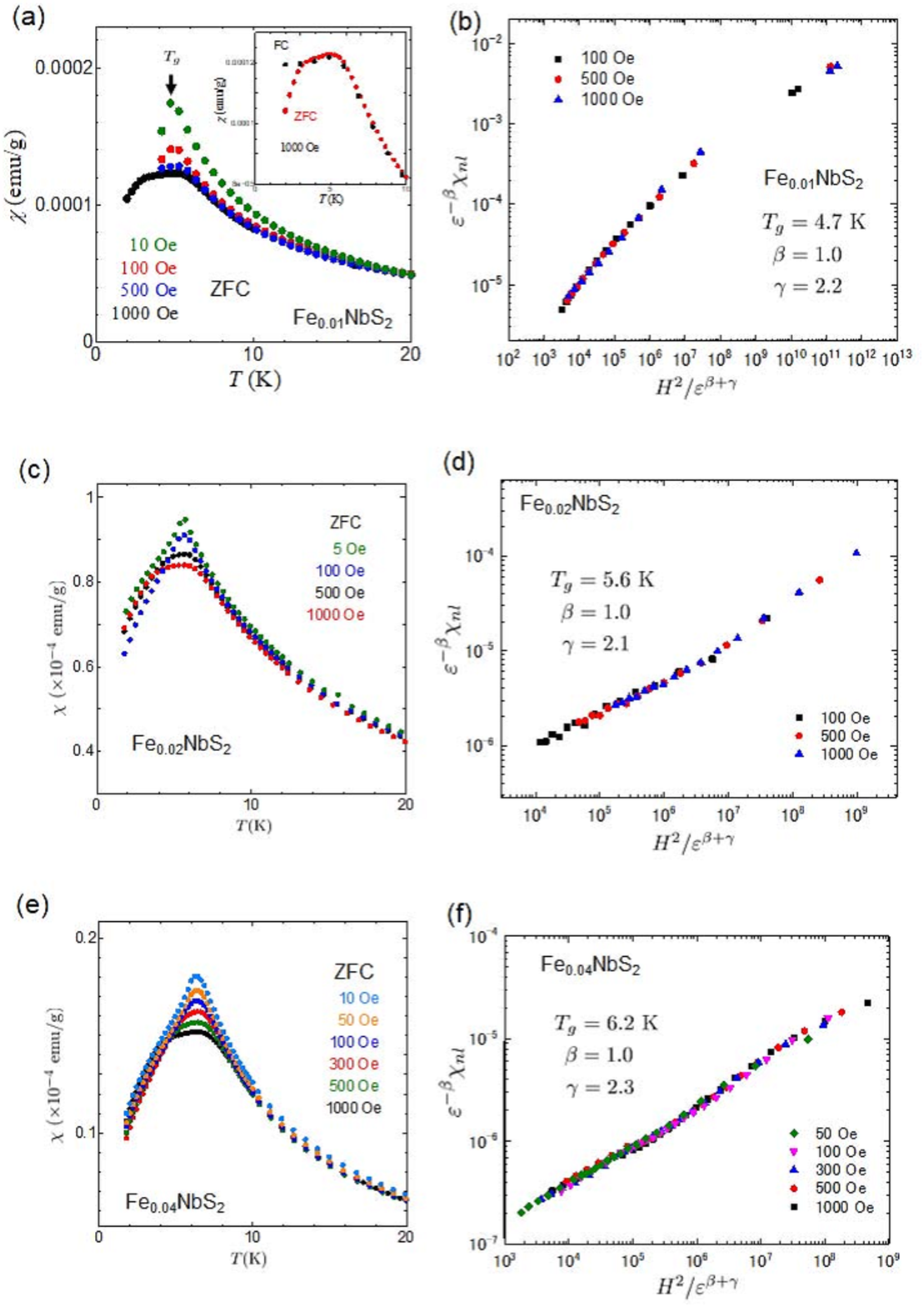}

\end{center}
\end{figure}

\newpage

\end{document}